\documentstyle[aps,epsfig]{revtex}

\newcommand{\cP}{{\cal P}}
\newcommand{\CovD}{D}
\newcommand{\Fs}[1]{\hat{{#1}}}
\newcommand{\fs}[1]{\hat{{#1}}}

\newcommand{\Klein}{\cP^2+m_c^2}
\newcommand{\Tr}{{\rm Tr}}
\newcommand{\tr}{{\rm tr}}

\newcommand{\sG}{\sigma G}
\newcommand{\be}{\begin{equation}}
\newcommand{\ee}{\end{equation}}
\newcommand{\bea}{\begin{eqnarray}}
\newcommand{\eea}{\end{eqnarray}}

\newcommand{\ba}{\begin{array}{l}}
\newcommand{\ea}{\end{array}}

\newcommand{\bb}{\begin{thebibliography}}
\newcommand{\eb}{\end{thebibliography}}
\newcommand{\ci}[1]{\cite{#1}}
\newcommand{\lab}[1]{\label{#1}}
\newcommand{\re}[1]{(\ref{#1})}
\newcommand{\Ds}{\displaystyle}

\newcommand{\Dirac}{\Fs{\cP}+im_c}

\begin{document}
\begin{center}
 {\bf
Current mass dependence of the quark condensate
and the constituent quark mass}\footnote{talk presented at the
International Symposium on Hadrons and Nuclei,
Seoul, February 20-22, 2001.}\\
{M. Musakhanov}\\
Theoretical  Physics Dept, Uzbekistan National University,\\
 Tashkent 7000174, Uzbekistan,
e-mail: yousuf@iaph.silk.org
\end{center}
\begin{abstract}
We  discuss  the
current mass dependence of the basic quantities of the quark models
-- constituent quark mass
$M$ and quark condensate $i<\psi^\dagger \psi >.$ The framework of the
consideration is
QCD instanton vacuum model.
\end{abstract}

\section{ Introduction}
Among the quark models, instanton vacuum based Effective Action approach
is a most promising since in this model the hadron properties and their
interactions features are closely related to the properties of QCD vacuum.

Without any doubts instantons are a very important component of the
QCD vacuum. Their properties are described by the average instanton size
$\rho$ and inter-instanton distance $R$. In 1982 Shuryak
\cite{Shu82} fixed them
phenomenologically as
\be
\rho \,=\, 1/3 \, fm,\, \, \, \, R\,=\, 1 \, fm.
\lab{rhoR}
\ee
From that time the validity of such parameters was confirmed by theoretical
variational calculations
\cite{DP84} and recent lattice simulations of the QCD vacuum (see
recent review \cite{latt}).
The presence of instantons in QCD vacuum very strongly affects light quark
properties, owing  consequent generation of
quark-quark interactions.
These effects lead to the formation of the massive constituent
interacting quarks. This implies spontaneous breaking of chiral symmetry
(SBCS), which leads to the collective massless excitations of the QCD
vacuum--pions. The most important degrees of freedom in low-energy QCD  are
these quasiparticles.  So instantons play a leading role in the formation of
the lightest hadrons and their interactions, while the confinement forces are
rather unimportant, probably.
The properties of the hadrons and their interactions
 are concentrated in the QCD Effective Action in terms
of quasiparticles.
The features of light quarks placed into instanton vacuum
are concentrated in the fermionic
determinant ${\det}_N$ (in the field of $N_+$ instantons and  $N_-$
antiinstantons) calculated by Lee and Bardeen(LB) in 1979 \cite{LB79}:
 \be
{\det}_N=\det B, \,\, B_{ij}=
im\delta_{ij} + a_{ji}, \lab{det_N}
\lab{detB}
\ee
and $a_{ij}$ is the overlapping matrix element of the quark zero-modes
$\Phi_{\pm , 0} $ generated by instantons(antiinstantons).
This matrix element is nonzero only between instantons and antiinstantons
(and vice versa) due to specific chiral properties of the zero-modes
and equal to
\be
a_{-+}=-<\Phi_{- , 0} | i\hat\partial |\Phi_{+ , 0} > .
\lab{a}
\ee
The overlapping of the quark zero-modes provides the propagating of the quarks
by jumping from one instanton to another  one.
So, the determinant of the infinite matrix
was reduced to the
determinant of the finite matrix in the space of \underline{only zero-modes}.
 From Eqs. \re{a}, \re{det_N}
it is clear that for  $N_{+}\neq N_{-}$
$
{\det}_N \sim m^{|N_{+}-N_{-}|}
$
which will strongly suppress the fluctuations of  $|N_{+}-N_{-}|.$
Therefore in final formulas we will assume $ N_{+}=N_{-}=N/2 .$

In \re{det_N}  we observe the competition
between current mass $m$ and overlapping matrix element $a \sim
\rho^{2}R^{-3}$.  With typical instanton sizes $\rho \sim 1/3 fm$ and
inter-instanton distances $R\sim 1fm$, $a$ is of the order of
the strange current quark mass, $m_s = 150 \, MeV$. So in this case it is very
important to take properly into account the current quark mass.

The fermionic determinant ${\det}_N$ averaged over
instanton/anti-instanton positions, orientations and sizes
can be considered as a partition function of light quarks $Z_N$.
Then the properties of the hadrons and their interactions are concentrated
in the QCD Effective Action. We calculate this one via
fermionic representation of ${\det}_N$ which provide
easy way for the averaging over instanton collective coordinates --
positions and orientations \cite{MK}.
This approach leads to the Diakonov-Petrov(DP)
Effective Action \cite{DPW} with a specific
choice of the degrees of freedom \cite{SM}.
It was shown that
DP Effective Action is a good tool in the chiral limit but failed
beyond this limit, checked by the calculations of
the axial-anomaly low energy theorems  \cite{SM}.
The solution of this puzzle related with the observation that
the fermionisation of ${\det}_N$ is not unique procedure and another
fermionic representation of ${\det}_N$  leads to a different choice
of the degrees of freedom in  the Effective Action.
Within this approach it was proposed so called
Improved Effective Action(IA) which is more properly taken into
account current quark masses  and satisfied
axial-anomaly low energy theorems also beyond the chiral limit \cite{M99}
at least at $O(m)$ order.

Completely another approach to the same problem was
developed by Pobylitsa \cite{Pobylitsa90}.
He directly summed up
planar diagrams for the propagator in the instanton medium in large
$N_c$ limit for two extreme cases: $N/VN_c -> 0$ and $N/VN_c -> \infty$.
We will compare his result for constituent mass with our one and will calculate
quark condensate within this approach too.

In the present case we concentrate on the calculation of the
current mass dependence of the quark condensate $i<\psi^\dagger \psi >$.
As a byproduct we find also current mass dependence of the
constituent quark mass $M$.
Since the quark condensate  does not dependent on the specific choice of the
degrees of freedom in  the Effective Action and entirely is defined
by  the current mass dependence of the partition function $Z_N$, it gives
important information on the accuracy of the fundamental LB result by
comparison with phenomenological data.  LB result \re{detB} itself has accuracy
which is $O(m^2 )$ order. We consider here the $m$ dependencies of
quark condensate $i<\psi^\dagger \psi >$ and  constituent quark mass $M$
within  DP and IA and  compare with results of
 slightly modified version of Improved Action(MIA),
 which has difference from IA on of $O(m^2)$ order terms.
So, both DP and IA approaches
must lead to the same $m$-dependence of the quark condensate
and must coincide with MIA approach, at least within $O(m^2)$ accuracy.
We consider first modification of Improved Action,
and further calculate the current mass dependencies of abovementioned quantities
within all variants of Effective Action.
The comparison of these results with the result of
the calculations within Pobylitsa approach provide independent test of the calculations.
Another test of the results is provided by heavy quark limit,
under the assumption that the gluon field strength $G_{\mu\nu}^a$ is
much less than the square of quark mass $m^2$.
We find that the quark condensate
in all approaches based on instanton vacuum model has almost the same rather strong
$m$ dependence and in the region $m\,>\,0.3\,\, GeV$
they are in accordance with heavy quark approximation.
As example, the strange quarks condensate
$<s^\dagger s>\sim 0.5<u^\dagger u>$
at $m_s \sim 0.15\,\, GeV.$
We find also rather strong $m$ dependence of the
constituent quark mass $M$. Since in Pobylitsa, IA and MIA cases
the total quark mass is $m+M$, the total mass
is almost constant in the region $m<0.2\,\, GeV$.
This dependencies contrasted very much naive expectations
and need detailed phenomenological analysis.

\section{  Modification of Improved Effective Action  }
First, accordingly Eqs. \re{detB} and \re{a} and
by introducing the Grassmanian
$(N_{+},N_{-})$ variables  $ \Omega_i$
$\bar\Omega_j$
we represent
\be \det B = \int d\Omega d\bar\Omega \exp (\bar\Omega B
\Omega ) , \ee
where
\be
\bar\Omega B \Omega  = \bar\Omega_i (im + a^{T})_{ij} \Omega_j =
-\Omega_j (\Phi^{+}_{j,0} (-i\hat\partial + im )_{ji}\Phi_{i,0})\bar\Omega_i
\lab{B}
\ee
This formula is transformed to:
\bea
\bar\Omega B \Omega
&=&
 \Omega_j (\Phi^{+}_{j,0}(i\hat\partial  (i\hat\partial + im )^{-1}
i\hat\partial + m^2 (i\hat\partial + im )^{-1})_{ji}\Phi_{i,0})\bar\Omega_i
\nonumber
\\
&=&
 \Omega_j (\Phi^{+}_{j,0}(i\hat\partial  (i\hat\partial + im )^{-1}
i\hat\partial )^{-1})_{ji}\Phi_{i,0})\bar\Omega_i + ...
\lab{B1}
\eea
where we are neglecting by $O(m^2 )$ term, since Lee-Bardeen result for
$det_N$ itself was derived within this accuracy.
The next step is to introduce   $N_{+},N_{-}$ sources
$\eta_i$ and $\bar\eta_j$
defined as:
\be
\bar\eta_{i}=-\Phi^{+}_{i , 0} \Omega_{i}i\hat\partial ,
\eta_{j} = i \hat\partial \Phi_{j , 0} \bar \Omega_{j}
\lab{eta}
\ee
Then $(\bar\Omega B \Omega)$ can be rewritten as
\be
(\bar\Omega B \Omega)\,=\,
- \bar\eta (i\hat\partial +im)^{-1}\eta =
- \sum_{ij}\bar\eta_{j}(i\hat\partial +im)^{-1}\eta_{i}
\ee
 and $\det B $ can be rewritten as
\bea
\det B &=& \int d\Omega d\bar\Omega \exp (\bar\Omega B \Omega )
\nonumber\\
&=&
\left(\det(i\hat\partial +im)\right)^{-1} \int d\Omega d\bar\Omega
D\psi D\psi^{\dagger}
\exp\int dx (\psi^{\dagger} (x)
(i\hat\partial\,+\,im)\psi (x)
\nonumber\\
&+& \sum_{i}(\bar\eta_{i} (x)\psi (x)
\,+\,\psi^{\dagger} (x)\eta_{i} (x)))
\eea
The integration  over Grassmanian variables $\Omega$ and $\bar\Omega$
(with the account  of the $N_{f}$ flavors
${\det}_N = \prod_{f}\det B_{f}$)
 provides finally the fermionized representation of
fermionic determinant \re{B1} in the form:
\be
\ba  \Ds
\det B = \int D\psi D\psi^{\dagger} \exp\left(\int d^4 x
\sum_{f}\psi_{f}^{\dagger}(i\hat\partial \,+\, im_{f})\psi_{f}\right)
    \\   \Ds
\times \prod_{f}\left\{\prod_{+}^{N_{+}} V_{+}[\psi_{f}^{\dagger} ,\psi_{f}]
\prod_{-}^{N_{-}}V_{-}[\psi_{f}^{\dagger},\psi_{f}]\right\}\; ,
\label{part-func}
\ea
\ee
where
\be
V_{\pm}[\psi_{f}^{\dagger} ,\psi_{f}]=
\int d^4 x \left(\psi_{f}^{\dagger} (x) i\hat\partial
\Phi_{\pm , 0} (x; \xi_{\pm})\right)
\int d^4 y
\left(\Phi_{\pm , 0} ^\dagger (y; \xi_{\pm} )
(i\hat\partial \psi_{f} (y)\right).
\lab{V}
\ee
Now the averaging over collective coordinates $\xi_\pm$ become trivial
problem.
The further steps are the exponentiation and the bosonization of
the integrand \cite{DPW}.
Finally, the corresponding partition function in terms of
constituent  quarks has a form \cite{M99}:
\be
Z_N
=\int d\lambda_{+} d\lambda_{-}
 D\Phi_{+}D\Phi_{-}
\exp\left(-S[\lambda_{+},\Phi_{+};\lambda_{-},\Phi_{-}]\right),
\label{Z}
\ee
where
\bea
S[\lambda_{+},\Phi_{+};\lambda_{-},\Phi_{-}] &=& -\sum_{\pm}
 \left( N_{\pm} \ln [\left(  \frac{4\pi^2
\rho^2}{ N_c} \right)^{N_f}\frac{N_{\pm}}{V\lambda_{\pm}}] - N_{\pm}\right)
+ S_{\Phi} + S_{\psi},
\lab{S}
\\
S_{\Phi} &=& \int d^4 x \sum_{\pm}
 (N_f - 1) \lambda_{\pm}^{-\frac{1}{N_f - 1}}
(\det\Phi_{\pm} )^{\frac{1}{N_f - 1}}   ,
\nonumber
\\
S_{\psi} &=& -
\Tr \ln ((-\hat k  + im_{f}\delta_{fg} +
i F(k_{1})F(k_{2}) \sum_{\pm}
\Phi_{\pm, fg }(k_{1}-k_{2})\frac{1\pm\gamma_{5}}{2})
 (- \hat k + i m_f )^{-1}).
\nonumber
\eea
Variation of the total action $S[\lambda_{+},\Phi_+;\lambda_{-}\Phi_-]$
over $\lambda_{\pm}$  leads to the saddle-point:
$$\lambda_{\pm} = (N_{\pm}^{-1} \int d^4 x
(\det\Phi_{\pm} )^{\frac{1}{N_f - 1}})^{(N_f -1)},$$
The additional variation over
$\Phi_{\pm}$ must vanish in the common saddle-point.
Since we take $N_{+} = N_{-} = N/2$, this one is
$$\Phi_{\pm ,fg}= \Phi_{\pm,fg}(0)=M_{f}\delta_{fg},$$
and
$$\lambda_{\pm}=
\lambda = \frac{2V}{N}\prod_{f} {M_f} ,$$
This condition leads to the  the saddle-point equation for the
momentum dependent constituent mass $M_{f}(k)$, i. e.,
\be
M_{f}(k)=M_{f}F^{2}(k).
\lab{M_f} \ee
The contribution of the quark loop to the saddle-point equation is
\be
\Tr \ln [ ( -\hat k + im + iF^{2}(k) \sum_{\pm} \Phi_{\pm}
\frac{1\pm\gamma_{5}}{2})( -\hat k + im )^{-1}].
\ee
Then we get the saddle-point equation
\be
N/V = 4N_c \int \frac{d^4 k}{(2\pi )^4}
\frac{ M_f F^{2}(k)(m_f + M_f F^{2}(k))}{k^2 + (m_f + M_f F^{2}(k) )^2}
\lab{saddle}
\ee
The form-factor $F(k)$ is related  to the zero--mode
wave function in momentum space $\Phi_\pm (k; \xi_{\pm}) $ \cite{DPW}.
We use  simplified expression  for this form-factor:
\be
F(k)=\frac{L^2}{L^2 + k^2},
\lab{F1}\ee
where $L^2 \sim 2/\rho^2 = 0.72 \, GeV^2 .$
which was proposed in \cite{Petal98}.

The important steps in the derivation of these formulas
\re{Z} and \re{S} are:\\
1. The fermionisation of \re{B1} (which is in fact not unique procedure);\\
2. Independent averaging over positions and orientations
of the instantons,
due to the small packing parameter of the instanton media --
$(\rho /R)^4 \sim (1/3)^4$;\\
3. The exponentiation and
the bosonization of the  partition function, described in \cite{DPW}.

The matrices $\Phi_{\pm}$, whose usual decomposition is
$
\Phi_{\pm} = \exp(\pm \frac{i}{2}\phi )M\sigma \exp(\pm \frac{i}{2}\phi ) ,
$
$\phi$ and $\sigma$ being $N_{f} \times N_{f}$
matrices, describes mesons
and $M_{fg}=M_f \delta_{fg}$.
At the saddle-point $\sigma = 1,\, \phi = 0$. The
usual decomposition for the  pseudoscalar fields
$\phi = \sum_{0}^{8}\lambda_{i}\phi_{i}$
may be used.
These mesons are considered as a small fluctuation near the saddle point.

The account of the fluctuations of number of instantons $N$
can be easily done.
Let
$N_\pm = 0.5(N \pm \Delta ), \,\,\,\,\,\, \Delta << N.$
Then assuming the saddle-points
$\Phi_{\pm fg }= \delta_{fg}M_{f\pm}$,
$ M_{f\pm} =M_{f}  (1 \pm \delta_f )$, $\delta_f << 1$ we find
additional to \re{saddle} another saddle-point equation
\be
\Delta /V = 4N_c \delta_f \int \frac{d^4 k}{(2\pi )^4}
\frac{m_f  M_f F^{2}(k) }{k^2 + (m_f + M_f F^{2}(k))^2}
\lab{saddle-delta}
\ee
Taking into account the definition of the condensate $i<\psi^\dagger \psi>$
\re{condensate1}
we find
\be
m_f \delta_f = \frac{\Delta}{V i<\psi_{f}^\dagger \psi_f >} (1 + O(m_{f}^2 ))
\ee
This formula leads to the $\Delta$--distribution,
which  is in accordance with general theorems \cite{DPW}.

\section{ Current mass dependence of constituent mass }
The saddle-point equation \re{saddle}  leads to the
momentum dependent constituent mass $M_{f}(k)=M_{f}F^{2}(k)$.
The constituent quark propagator, as it is follows from \re{Z},
has a form:
\be
 S \,=\,(-\hat k \,+ \,i(m_{f} \,+ \,  M_{f}F^{2}(k) ))^{-1},
 \lab{propagator}
\ee
In IA approach analogous  saddle-point condition:
 \be
N/V = 4N_c \int \frac{d^4 k}{(2\pi )^4}
\frac{ M^{IA}_{f} F^{2}(k)(1+m_{f}^{2}/k^{2})(m_f + M^{IA}_{f} F^{2}(k)(1+m_{f}^{2}/k^{2}))}{k^2 + (m_f +
M^{IA}_{f} F^{2}(k)(1+m_{f}^{2}/k^{2}) )^2}
\lab{saddle-IA}
\ee
 also define $M^{IA}_{f}$
and the constituent quark propagator has a form:
\be
 S^{IA} \,=\,(-\hat k \,+ \,i(m_{f} \,+ \,  M^{IA}_{f}F^{2}(k)(1+m_{f}^{2}/k^{2}) ))^{-1},
 \lab{propagator-IA}
\ee
On the other hand, DP Effective Action leads to the propagator:
\be
 S^{DP} \,=\,(-\hat k \,+ \,  M^{DP}_{f}(k) ))^{-1},
 \lab{propagator-DP}
\ee
where $M^{DP}_{f}(k)=M^{DP}_{f} F^{2}(k)$
is followed from analogous saddle-point equation:
\bea
\frac{4VN_{c}}{N}\int \frac{d^4 k}{(2\pi)^{4}}\frac
{M^{DP\, 2}_{f}(k) }
{k^{2} + M_{f}^{DP\, 2}(k)}
= 1 \,-\, \frac{M^{DP}_{f}m_{f} V N_{c} }{2\pi^2 \rho^2},
\lab{saddle-DP}
\eea
As was mentioned in the Introduction, Pobylitsa \cite{Pobylitsa90}
in quenched approximation
directly summed up
planar diagrams for the propagator in the instanton medium in large
$N_c$ limit for two extreme cases: $N/VN_c -> 0$ and $N/VN_c -> \infty$.
In the first (and most interesting) case his result can be summarized in the form:
\be
 S_P \,=\,(-\hat k \,+ \,i(m \,+ \,  M^{P}(k) ))^{-1},
 \lab{propagator-P}
\ee
where
\bea
M^P (k)&=& M_0 F^{2}(k) [(1+m^{2}/d^{2})^{1/2} - m/d],
\nonumber
\\
d &=& \left(\frac{0.08385}{2N_c} \right)^{0.5} \frac{8 \pi \rho }{R^2 } =
0.198\, GeV .
\lab{M^P}
\eea
\begin{figure}[hbt]
   \epsfysize=8cm
   \epsfxsize=10cm
   \centerline{\epsffile{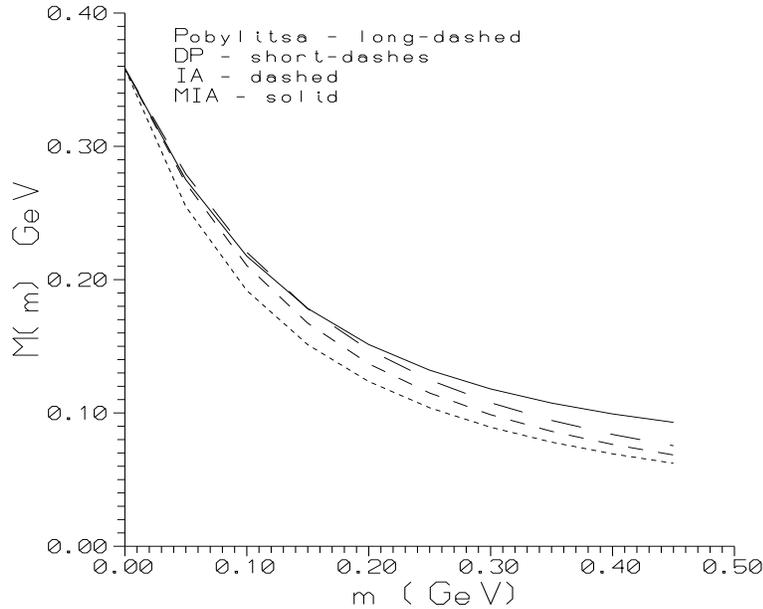}}
\vskip 0.5cm
{\caption{\label{fig1}\em
Current mass dependence of the constituent mass:
Solid line -- Modified Improved Action calculations.
Long dashed line -- Pobylitsa approach calculations.
Dashed line -- Improved Action calculations.
Short dashed line -- DP Action calculations.}}
\end{figure}
Fig.1 represent different versions of the current mass dependence of
constituent mass derived from saddle-point equations \re{saddle}, \re{saddle-IA}, \re{saddle-DP}
and Pobylitsa result \re{M^P}.

\section{ Current mass dependence of the quark condensate}

First, we calculate the quark condensate by using the evident formula
\bea
i<\psi_{f}^{\dagger}\psi_{f} >\,&=&\,
V^{-1}Z^{-1}_{N}\frac{\partial Z_{N}}{\partial m_{f}}
\nonumber\\
&=& \Tr i[(-\hat k + im_{f} +i M_{f}(k) )^{-1} - (-\hat k +
im_{f})^{-1}].
\lab{condensate1}\eea
In \re{condensate1}
the saddle-point condition was taken into account.
It is evident, the condensates in IA and Pobylitsa approaches
are calculated with similar formula.

With DP Action  \cite{DPW} the condensate is
\bea
i<\psi_{f}^{\dagger}\psi_{f} >^{DPW}
\,=\,\frac{N_c M_{f}^{DP}}{2\pi^2 \rho^2}.
\lab{condensate3}
\eea
Simple numerical calculations  leads to the condensate as a function of
current mass presented in Fig.2.
We present here also heavy quark approximation result
for the quark condensate.
In this limit, for heavy quarks  we have to use the expansion over
small parameter $G /m^2$
under the assumption that the gluon field strength $G_{\mu\nu}^a$ is
much less than the square of quark mass $m^2$ \cite{VZNS83,AMT}. Then
\bea
\int d^{4}x
i< \psi^\dagger \psi>&=&\Tr\left(\frac{i}{\Dirac}
-\frac{i}{\fs{p}+im}\right)
=m\Tr\left(\frac{1}{\Klein+\frac{g}{2}\sG}
-\frac{1}{p^{2}+m^{2}}\right)\nonumber\\
&=&
\int d^{4}x (
\frac{g^{2}}{24\pi^{2}m}\tr_{c}
G_{\alpha\beta}^{2} +
\frac{ig^{3}}{360\pi^{2}m^3}\tr_{c}
G_{\alpha\beta}G_{\alpha\gamma}G_{\beta\gamma}  + ...)
\lab{heavy}
\eea
Here
$\sG\equiv\sigma_{\mu\nu}G_{\mu\nu}$,
$\sigma_{\mu\nu}\equiv\frac{i}{2}[\gamma_\mu , \gamma_\nu]$,
$\Fs{\cP}\equiv\cP_\mu \gamma_\mu$
and $\cP_\mu=i\CovD_\mu=i(\partial_\mu-igA_\mu^a t^a)$.

In the instanton vacuum:
$$ \int d^{4}x \tr_{c}g^{2} G_{\alpha\beta}^{2} =  (4\pi )^2 N, \,\,\,\,
\int d^{4}x \tr_{c}g^{3} G_{\alpha\beta}G_{\alpha\gamma}G_{\beta\gamma}
= - i (4\pi )^2 \frac{6}{5\rho^2 }N $$
and
\bea
i< \psi^\dagger \psi> = \frac{ 2  }{3 m R^4} +
\frac{4 }{75 \rho^2 m^3 R^4} =
\frac{ 2  }{3 m R^4}(1 + \frac{12}{150\rho^2 m^2})
\lab{heavy-m}
\eea
\begin{figure}[hbt]
   \epsfysize=8cm
   \epsfxsize=10cm
   \centerline{\epsffile{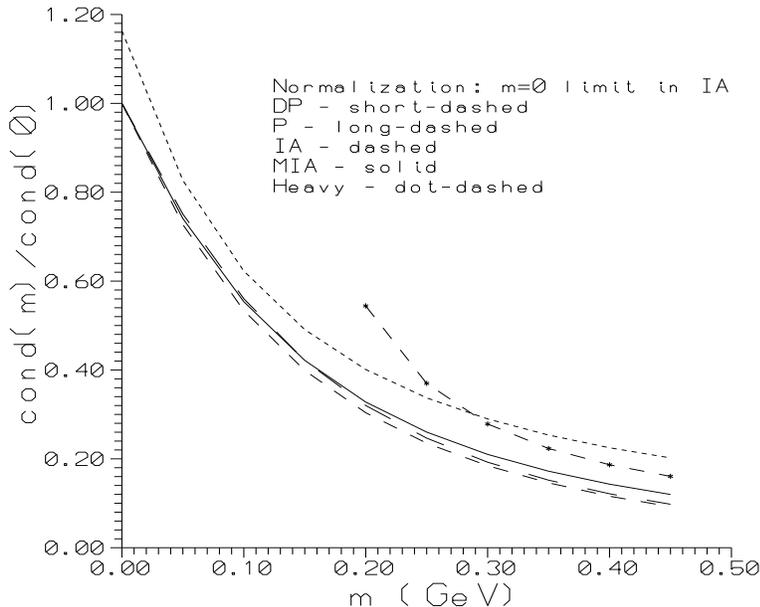}}
\vskip 0.5cm
{\caption{\label{fig2}\em
Current mass dependence of the quark condensate:
Solid line -- Modified Improved Action calculations.
Long dashed line -- Pobylitsa approach calculations.
Dashed line -- Improved Action calculations.
Short dashed line -- DP Action calculations.
Dashed-dot line -- Heavy quark approximation \protect{\re{heavy-m}}.}}
\end{figure}
We see that MIA,IA, DP and Pobylitsa results almost
coincide with each other and in
the good correspondence with
heavy quark approximation at $m\, >\, 0.3 \, GeV$.

\section{  Conclusion }
It were investigated the current quark mass dependencies of
quark condensate and constituent quark  mass
in QCD instanton vacuum model. It were considered different approaches:\\
1. Diakonov\&Petrov
effective action (see recent papers \ci{DPW});\\
2. Improved Action \ci{M99}
and presented here Modified Improved Action.\\
They are essentially based on Lee\& Bardeen result for the quark determinant
in instanton vacuum background \ci{LB79}.\\
3. Direct summation of
planar diagrams for the propagator in the instanton medium  in quenched
approximation \ci{Pobylitsa90}.\\
All of these approaches leads to rather fast dependencies of
quark condensate and constituent quark  mass on current quark mass, as were
demonstrated in Figs 1,2.
Then, the strange quarks condensate
$<s^\dagger s>\sim 0.5<u^\dagger u>$
at $m_s \sim 0.15\,\, GeV$ and
total quark mass  $m+M$
is almost constant in the region $m<0.2\,\, GeV$.

We conclude that strange quark physics
might be very different from usual expectation based on old-fashioned
quark model and demand careful phenomenological reanalysis.
\vskip .5true cm
{\bf Acknowledgments}
\vskip .5true cm
I am very grateful to M.Birse, D.Diakonov, V.Petrov, P.Pobylitsa and
M.Polyakov for useful discussions.

\end{document}